# Cumulative labelling with thymidine analogues when the steady–state assumption is violated

**Darragh M. Walsh**[1, 2]

[1] Insight Centre for Data Analytics, Dublin City University, Ireland.
[2] School of Mathematics and Statistics, University of Melbourne, Australia.

Correspondence: darragh.walsh@dcu.ie

**Abstract**

We present modelling results that examine the consequences of implementing cumulative labelling with thymidine analogues, to estimate the cell cycle time and growth fraction of dividing cells, when the steady-state assumption is violated. We fix the value of the cell cycle time a priori and examine whether cumulative labelling can reproduce this value. We find that the cumulative labelling technique systematically overestimates the growth fraction and cell cycle time in non-steady cell populations. Our results suggest an explanation for discrepancies in experimental measurements of oligodendrocyte precursor cell properties using cumulative labelling. These results also emphasise the utility of using computational models to determine what violating the assumptions of experimental techniques would look like in the laboratory before experiments are undertaken.

**Introduction**

The cumulative Bromodeoxyuridine (BrdU) labelling (CL) technique was introduced by Nowakowski et al.[1] to measure the time it takes for a dividing cell to complete the cell cycle, $T_c$. A BrdU injection at time $t_1$ labels *all* the cells in S-phase with BrdU. At a later time, some of those cells will have left S-phase whilst other cells will have entered S-phase. A further injection will label all cells in S-phase at time $t_2$. Since labelled cells remain positive for BrdU after they have left S-phase, periodic injections should eventually cumulatively label all dividing cells.

Nowakowski and colleagues noted that the CL technique is an appropriate method to estimate the length of the cell-cycle, $T_c$, and the percentage of cells that are dividing, the growth fraction GF, if 1) the cells are part of a single asynchronous population and 2) the cell population is growing at a steady state. The steady-state assumption is met, for example, in situations where there is a clear anatomical separation between parent cells and daughter cells which migrate away from the anatomical region of interest.

In a later work, Hayes et al.[2] widened the scope of this method by demonstrating the benefits of using two markers, BrdU and tritiated thymidine ($^3$H-TdR), simultaneously to estimate the number of cells entering and leaving S-phase in the developing mouse cortex. This double labelling technique has been shown to be successful in numerous studies, notably in combination with the protein Ki67 (which labels cells in all stages of the cell cycle except $G_0$)[3]. However, it is clear that the CL technique with thymidine analogues such as BrdU remains influential, though the technique's explicit assumptions are not always met.

In this work we present a computational model that examines the effect of implementing the CL technique when the steady–state assumption is violated. We compare our modelling results to established estimates of $T_c$ and GF for oligodendrocyte precursor cells in the mouse brain using the CL technique [4] and more recent estimates of the GF from expression of Ki 67 [5], also from the mouse brain.

Our model simultaneously tracks dividing cell identities in two scenarios: when the measurement technique is assumed to be perfect and when the measurement technique is imperfect (not sensitive to changes in cell fate i.e. when the progeny of a division become quiescent but retain their BrdU labelling). Further details may be found in the online Methods section.

**Results**

In all the following scenarios, we assume that $T_s$ and $T_c$ are constant (15 hours and 45 hours respectively) and examine whether non-steady–state cell population scenarios can recover these values. Initially, we assume there are 1000 quiescent cells (cells in cell cycle stage $G_0$) and 1000 dividing cells (cells that are in any other active stage of the cell cycle), and zero differentiated cells. Hence GF = 0.5 initially. We retain the assumption that dividing cells are asynchronously distributed throughout the cell cycle. The labelling index curve LI(t) represents the proportion of cells in the population that are positive for BrdU at successive periodic exposures to BrdU. When all cells are labelled, the slope of LI(t) is zero. The value of LI(t) at which this occurs is the GF. This time at which this occurs is $T_c - T_s$.

In Fig. 1a we display the results of a computational model where half of the progeny of a population of dividing cells leave the anatomical region completely (they are 'cleared') whilst the other half remain to re-enter the cell cycle. In this scenario the steady–state assumption of the CL technique is valid and the x-axis time at which this saturation plateau is reached is $T_c - T_s$ = 30 hours, as required. The labelling index LI(t) (red curve) increases linearly until it reaches a plateau saturation value where all proliferating cells are labelled. The blue curve represents the actual (perfect measurement) GF, which overlaps with the LI curve when all cells are labelled. This scenario models the correct application of the CL technique, with all assumptions met. (In the Supplementary Material Fig. S1, we describe how this linearity can be apparently preserved even when the steady state assumption is violated).

However, the population of quiescent cells may not be entirely quiescent. For example, in a cell population dominated by homeostasis, proliferation may be driven by local cues to ensure the maintenance of the density of the cell population[6]. In Fig. 1b–e, we violate the steady-state assumption by dropping the requirement that one cell from each division clears. Instead, for illustration, we assume 80% of the proliferating cells become quiescent after mitosis, with the remaining 20% free to re-enter the cell cycle.

The CL technique is not suitable in these scenarios since it will over-estimate the GF by failing to distinguish between labelled quiescent and labelled actively dividing cells. Thus, we see in Fig. 1b that the actual GF (blue curve) is decreasing whilst the CL technique measured LI (red curve), whose eventual plateau defines the GF, is increasing until the plateau saturation is reached at approximately 0.63. Interpreting a CL experiment displaying these cell counts would lead to the erroneous conclusion that $T_c - T_s$ is equal to approximately 40 hours, instead of the pre-determined value of 30 hours. In Fig. 1c-d, we add biologically realistic mechanisms to the model: the differentiation of a small percentage of quiescent cells (Fig. 1c) and the gradual dilution of label fluorescence (Fig, 1d) after multiple divisions. Differentiation of quiescent cells removes cells from the denominator in the GF, so keeping all other quantities fixed we would expect to see the GF increase, as observed in Fig. 1c. Since the CL technique is designed to reach a plateau when all dividing cells are labelled, we expect label dilution to extend the time interval $T_c - T_s$, as observed in Fig. 1d.

Finally, in Fig. 1e, the cell differentiation and BrdU label dilution processes are both present. Measurement by CL fails to capture the pre-determined $T_c$, instead estimating $T_c - T_s$ to be approximately 60 hours, and CL estimates GF to be close to 1, though the true actual value is 0.2.

**Discussion**

The use of the CL technique when the steady–state assumption is likely to be invalid (for example in the corpus callosum or cortex), possibly combined with cell differentiation (which increases GF) and label dilution after several divisions (which lengthens $T_c$), may explain the discrepancies between the estimates of $T_c$ and GF in Young et al.[4] (Figure 1 & 2 of that work) and Spitzer et al.[5] (Figure S2 of that work) in the postnatal mouse corpus callosum and cortex.

Young et al.[4] used the CL technique and found that essentially all oligodendrocyte progenitor cells (OPCs) were active in the cell cycle (GF ≈ 1), with long cell cycle times, in several regions of the mouse central nervous system. We propose that the measurements of Young et al. are qualitatively similar to the red curves in Fig 1b-e, whilst the measurements of Spitzer et al. are qualitatively similarly to the blue curves in Fig 1b-e.

Spitzer and colleagues used FUCCI and Ki 67 cell labelling methods to estimate the proportion of OPCs in each stage of the cell cycle and the GF. They observed a much smaller GF (that declined with age to less than 0.05 before adulthood in both the cortex and the corpus callosum), which is incompatible with the CL measurements of GF and $T_c$ by Young et al. (since measurement of $T_c$ and GF are coupled in the CL technique, $T_c$ = GF / $m$, where $m$ is the slope of LI). The live imaging analysis of OPC homeostasis in the adult mouse brain by Hughes et al.[6] also favours shorter cell cycle times, consistent with Spitzer et al., over the much longer cell cycle times reported by Young et al. A small and declining OPC GF with age in the postnatal mouse corpus callosum was also reported in Walsh et al.[7], where Ki 67 was used to distinguish actively dividing cells.

Given the importance of accurate measurements of the GF and $T_c$ of OPCs when assessing interventions to curb demyelinating diseases such as multiple sclerosis, and the continued debate[8] surrounding the interpretation of BrdU labelling to signal human neurogenesis (co-expressed with a neuronal marker such as NeuN), we hope this work will lead to a careful analysis of what violation of the assumptions of experimental techniques would look like in laboratory measurements. Computational modelling should be considered to provide conceptual support and analyse predictions in biological scenarios where verifying that a technique's assumptions are met may be challenging in practice.

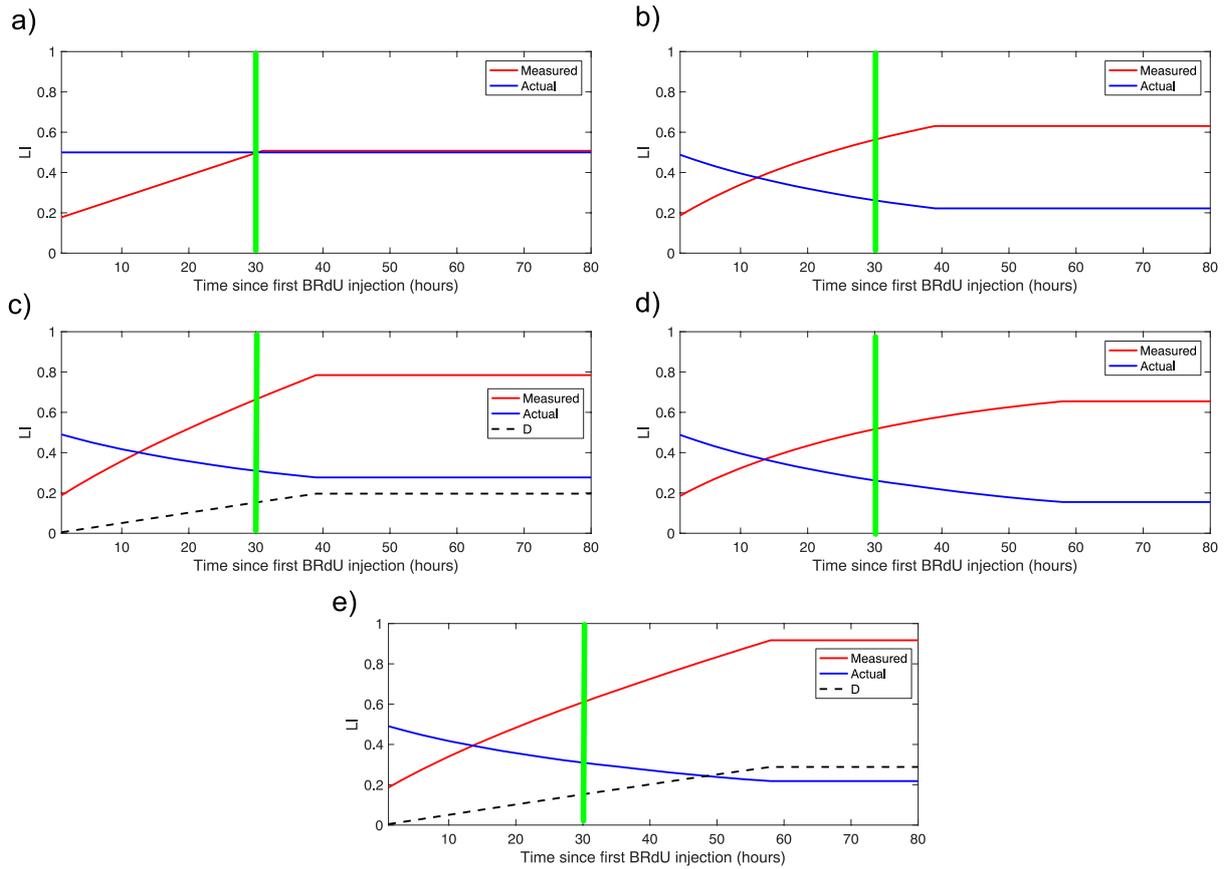

**Figure 1. A computational model of CL when the steady state assumption is invalid.** In each case, the values of $T_s$ and $T_c$ are predetermined, with $T_s$ = 15 hours and $T_c$ = 45 hours. We examine whether a model of CL can capture these values when the steady – state assumption is violated. The green vertical line defines the *x*-axis location of the actual value of $T_c - T_s$ in each scenario (30 hours).

a) The labelling index LI (cumulative labelling of cells with BrdU, red curves) ceases when all dividing cells in the population have passed through S-phase of the cell cycle. The resulting saturation plateau (red horizontal line) defines the growth fraction GF. Since the length of $T_s$ is 15 hours, the red/blue line represents a $T_c$ of 45 hours ($T_c - T_s$ = 30), as required.

b) Violating the steady-state assumption (in this instance by allowing 80 % of post-mitotic cells to become quiescent) results in an overestimation of GF. The red line is the proportion of labelled dividing cells, whether actively dividing or quiescent, that would be measured by the CL technique, whilst the blue line is the actual proportion of labelled actively dividing cells. The apparent value of $T_c - T_s$ is 40 hours.

c) With 80% of progeny entering $G_0$ as in b), we force 1% of quiescent cells to differentiate and remove them from the dividing population per (2 hour) time-step. The apparent GF is inevitably increased (red curve). The D curve denotes the proportion differentiated cells in the entire cell population. The apparent value of $T_c - T_s$ is 40 hours.

d) With 80% of progeny entering $G_0$ as in b), we force 1% of actually dividing labelled cells to transition to unlabelled dividing cells (label dilution) per (2 hour) time-step. The apparent $T_c$ increases to approximately 75 hours ($T_c - T_s$ = 60 hours) and the apparent GF is driven towards unity (red line) while the actual GF (blue curve) is driven to less than 0.2 at saturation.

e) Combining violation of the steady-state assumption with 1% differentiation and 1% dilution yields an apparent GF of close to 1 and an apparent measured $T_c$ of 75 hours ($T_c - T_s$ = 60 hours). Such mechanisms may explain the observations by Young et al.[4] of the GF and $T_c$ in several regions of the mouse central nervous system using the cumulative labelling technique with thymidine analogues.

**Availability of data and material**
All computer code required to generate the results are available from the author.

**Competing interests**

There are no competing interests.

**Funding**

This work was funded by fellowships awarded by the University of Melbourne and the AIB Chair in Data Analytics at Dublin City University. I am also grateful to Barry Hughes for funding from Australian Research Council Discovery grant (DP140100339).

**Acknowledgements**

I am grateful to David Gonsalvez, Georgina Craig and Barry Hughes for helpful discussions.

## Methods

The deterministic computational model that generated Fig. 1 was implemented using MATLAB and is available by request from the author. The parameters $T_s$ and $T_c$ used in generating Fig. 1 are for illustrative purposes and do not correspond to particular experimental measurements. A collection of parameters relevant to modelling OPCs generally are presented in Walsh et al.[8].

We begin with 1000 dividing cells, 1000 quiescent cells and zero differentiated cells, so that the GF is initially 0.5. We choose constant values of $T_c = 45$ hours and $T_s = T_c / 3$ throughout and measure the ability of the CL technique to capture this value of $T_c$ and the (evolving) GF. We assume that new labelling events occur discretely every 2 hours (the time-step of the algorithm corresponding to periodic exposure to BrdU) and that the number of cells that enter S–phase every two hours (after the initial injection) is equal to:

(the total actual number of actively dividing cells) $\times$ (2 / $T_c$).

We assume that all cells are part of a single asynchronous population (the first assumption of Nowakowski et al.[1]). The 'Measured' red curves in Fig. 1 refer to a cell counting mechanism that is not sensitive to the distinction between labelled actively dividing cells and labelled cells that are quiescent (having previously been through S-phase and exposed to BrdU).

To generate an actual GF that is qualitatively similar to recent measurements (low and decreasing with time)[5], in Fig. 1 we assumed that 80% of progeny became quiescent cells, whilst the other 20% re-entered the cell cycle. The effect of other proportions may be seen by changing the 'CellFate' variable in the computer program from 0.2 to any value in the range [0, 1].

The labelling mechanism/algorithm stops when there are no unlabelled dividing cells left.

If cell differentiation is present, we assume that only the initial population of quiescent cells can differentiate, which greatly simplifies the algorithm without any qualitative difference in the results. We assume that there is no difficulty classifying differentiated cells (even though they may still be BrdU labelled).

If label dilution is present, we assume that a fixed percentage (1% in Fig. 1d) of the actual labelled dividing cells lose their labelled identity per time–step.

Opening the Driver_CL.m file in MATLAB and clicking 'Run' will generate Fig. 1 and Fig S1. Further instruction may be found in the Readme.doc file.

**Supplementary Material**

A key feature of the CL technique is that when the steady state assumption is valid, we see a linear increase in the proportion of labelled dividing cells. Under the reasonable assumptions of our modelling, apparent linearity may also be realised when the steady state assumption is invalid, illustrated in Fig S1.

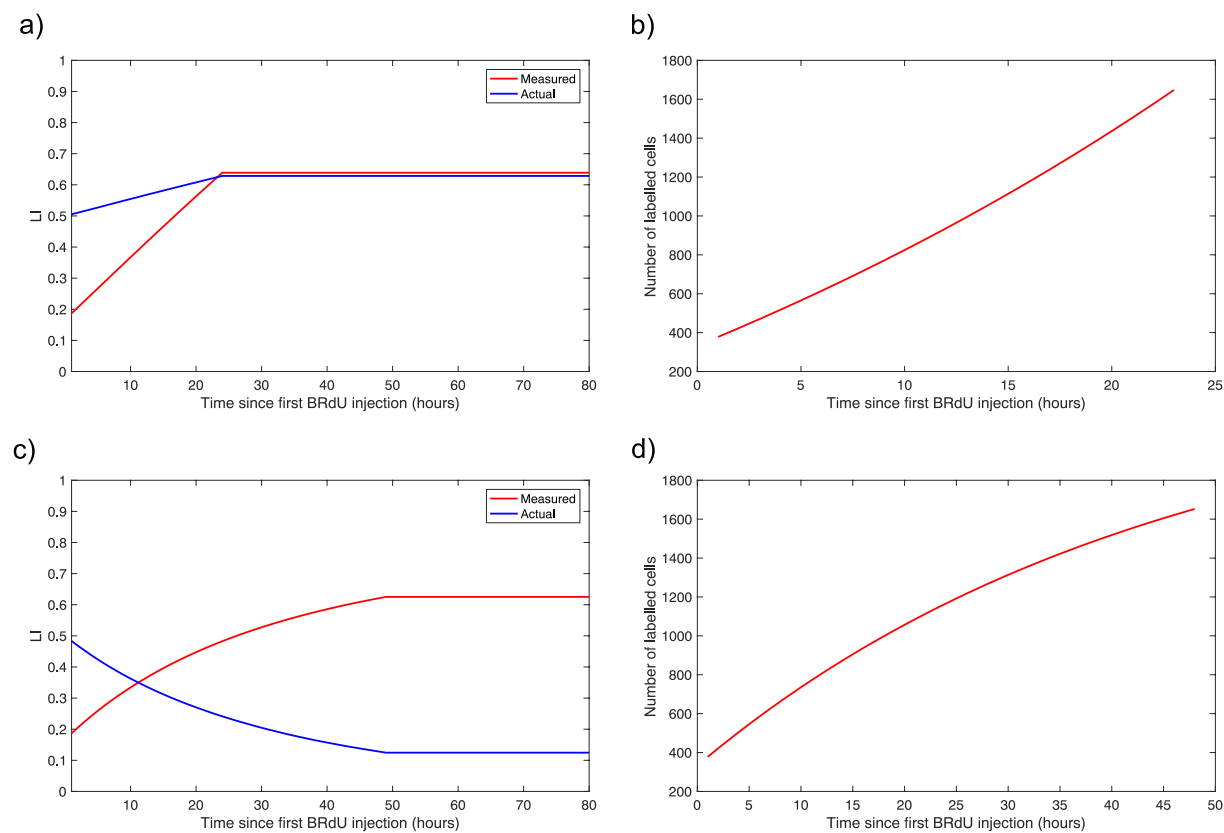

**Figure S1. The linearity of the labelling index curve.** a) An apparently linear labelling index curve LI is generated from a cell-cycle model where all post–mitotic cells re–enter the active cell cycle (the algorithm input parameter CellFate = 1). b) However, the number of measured labelled cells in this scenario increases exponentially.
c) The LI of a model where all post-mitotic cells become quiescent. D) Note that the number of measured labelled cells still increases despite no cells re-entering the cell cycle (CellFate = 0, so the actual number of dividing cells remains at the level after the first injection of BrdU). The apparent increase in the number of labelled cells is due to quiescent cells being erroneously counted as labelled dividing cells. Both models used $T_c$ = 45 hours and $T_s$ = 15 hours, as in Fig. 1.